\documentclass[sigconf]{acmart}

\usepackage{tikz}
\usepackage{mdframed}
\usepackage[normalem]{ulem}

\usepackage{booktabs}
\usepackage{tabularx}
\usepackage{array}
\usepackage{siunitx}

\usepackage[most]{tcolorbox}
\tcbuselibrary{listings,breakable}

\usepackage{listings}
\usepackage{pifont}
\usepackage{enumitem}
\usepackage{ragged2e}
\usepackage{soul}

\setcopyright{none}
\renewcommand\footnotetextcopyrightpermission[1]{}

\newcommand{\cmark}{\ding{51}}
\newcommand{\xmark}{\ding{55}}
\newcommand{\model}{ClarifyCodeBench}

\lstdefinestyle{promptstyle}{
  basicstyle=\ttfamily\footnotesize,
  columns=fullflexible,
  breaklines=true,
  breakatwhitespace=false,
  keepspaces=true,
  showstringspaces=false,
  upquote=true
}

\newtcblisting{promptbox}[1][]{%
  listing only,
  breakable,
  enhanced,
  colback=black!2,
  colframe=black!35,
  arc=2mm,
  boxrule=0.4pt,
  left=1.5mm,
  right=1.5mm,
  top=1.2mm,
  bottom=1.2mm,
  fonttitle=\bfseries,
  listing options={style=promptstyle},
  #1
}

\newtcolorbox{findingbox}{
  colback=white,
  colframe=black,
  boxrule=0.8pt,
  arc=3mm,
  left=6pt,
  right=6pt,
  top=6pt,
  bottom=6pt,
  width=\linewidth
}

\newcommand{\finding}[2]{
  \begin{findingbox}
  \noindent\textbf{Findings #1:}\quad #2
  \end{findingbox}
}

\definecolor{spgray}{RGB}{235,235,235}
\definecolor{spblue}{RGB}{187,206,240}
\definecolor{sporange}{RGB}{242,209,183}
\definecolor{spgreen}{RGB}{210,227,202}
\definecolor{spblueb}{RGB}{210,223,239}

\newcommand{\PromptBlock}[3]{%
\begin{tcolorbox}[
  enhanced,
  colback=#1,
  colframe=#1,
  boxrule=0pt,
  arc=0mm,
  left=2.5mm,
  right=28mm,
  top=1.6mm,
  bottom=1.6mm,
  before skip=0mm,
  after skip=0mm,
  overlay={
    \node[
      anchor=north east,
      font=\bfseries\itshape\footnotesize,
      text=black!65,
      align=right
    ] at ([xshift=-2mm,yshift=-1.6mm]frame.north east) {#3};
  }
]
\small\RaggedRight #2
\end{tcolorbox}%
}

\definecolor{casegray}{RGB}{235,235,235}
\definecolor{casegreen}{RGB}{214,234,214}
\definecolor{casered}{RGB}{242,220,220}
\definecolor{casegrayhead}{RGB}{223,223,223}
\definecolor{casegreenhead}{RGB}{196,221,196}
\definecolor{caseredhead}{RGB}{233,204,204}
\definecolor{casehighlight}{RGB}{255,245,170}

\sethlcolor{casehighlight}

\lstdefinestyle{casepython}{
  language=Python,
  basicstyle=\ttfamily\scriptsize,
  keywordstyle=\normalfont,
  commentstyle=\itshape,
  showstringspaces=false,
  columns=fullflexible,
  keepspaces=true,
  breaklines=true,
  breakatwhitespace=false,
  frame=none,
  aboveskip=0pt,
  belowskip=0pt
}

\newtcolorbox{reqbox}{
  enhanced,
  width=\linewidth,
  colback=casegray,
  colframe=casegray,
  colbacktitle=casegrayhead,
  coltitle=black!65,
  fonttitle=\itshape\footnotesize,
  title=Requirement,
  boxrule=0pt,
  arc=0mm,
  boxsep=0mm,
  left=2mm,
  right=2mm,
  top=1.2mm,
  bottom=1.2mm,
  toptitle=1mm,
  bottomtitle=1mm,
  lefttitle=2mm,
  righttitle=2mm,
  before skip=0mm,
  after skip=0mm
}

\newtcolorbox{codebox}[3][]{
  enhanced,
  width=\linewidth,
  colback=#3,
  colframe=#3,
  colbacktitle=#2,
  coltitle=black!65,
  fonttitle=\itshape\footnotesize,
  title=#1,
  boxrule=0pt,
  arc=0mm,
  boxsep=0mm,
  left=2mm,
  right=2mm,
  top=1.2mm,
  bottom=1.2mm,
  toptitle=1mm,
  bottomtitle=1mm,
  lefttitle=2mm,
  righttitle=2mm,
  before skip=0mm,
  after skip=0mm,
  valign=top
}

\begin{document}

\title[ClarifyCodeBench]{
ClarifyCodeBench: Evaluating LLMs on Clarifying Ambiguous Requirements for Code Generation
}


\author{Zheng Fang}
\affiliation{%
  \institution{Peking University}
  \city{Beijing}
  \country{China}
}
\email{fangz@pku.edu.cn}

\author{Dongming Jin}
\affiliation{%
  \institution{Peking University}
  \city{Beijing}
  \country{China}
}
\email{dmjin@stu.pku.edu.cn}

\author{Yihong Dong}
\affiliation{%
  \institution{Peking University}
  \city{Beijing}
  \country{China}
}
\email{dongyh@stu.pku.edu.cn}

\author{Yongmin Li}
\affiliation{%
  \institution{Peking University}
  \city{Beijing}
  \country{China}
}
\email{liyongmin@pku.edu.cn}

\author{Kechi Zhang}
\affiliation{%
  \institution{Peking University}
  \city{Beijing}
  \country{China}
}
\email{zhangkechi@pku.edu.cn}

\author{Zhi Jin}
\affiliation{%
  \institution{Peking University}
  \city{Beijing}
  \country{China}
}
\email{zhijin@pku.edu.cn}

\author{Ge Li}
\authornote{Corresponding author.}
\affiliation{%
  \institution{Peking University}
  \city{Beijing}
  \country{China}
}
\email{lige@pku.edu.cn}

\begin{abstract}
Large Language Models (LLMs) have emerged as powerful programming assistants. However, the efficacy of code generation is fundamentally constrained by the quality of input requirements, which, in real-world software development, are frequently ambiguous, incomplete, or underspecified. While LLMs excel at one-shot code synthesis, their ability to proactively clarify intent remains underexplored, as a critical trait for robust software engineering. Existing benchmarks largely overlook this interactive bottleneck, assuming perfectly specified prompts that do not reflect the iterative nature of requirement elicitation.
To bridge this gap, we introduce \model, a novel interactive benchmark specifically designed to evaluate LLMs' capability in resolving requirement ambiguity. Constructed from real-world programming tasks, \model{} features high-quality manual annotations, including $N$ unique ambiguity types, associated clarification questions, and corresponding ground-truth answers. Furthermore, we formalize two rigorous metrics to assess the interaction quality: Turn-discounted Key Question Rate (TKQR), which penalizes inefficient questioning, and Optimal Round Adherence (ORA), which measures the precision of the elicitation process. We conduct a systematic evaluation of six state-of-the-art LLMs using \model{}. Our empirical results yield three critical insights: 1) Capability Decoupling: Strong code generation performance does not inherently translate to effective requirement clarification; 2) The Reasoning Paradox: While increased computational "thinking" (e.g., via reasoning models) enhances code correctness, it yields marginal gains in identifying ambiguities; 3) The Multi-ambiguity Ceiling: LLMs' clarification performance degrades sharply as the density of ambiguities increases, revealing a significant bottleneck in handling complex, real-world specifications. Our work underscores the necessity for future AI4SE research to transition from static synthesis to interactive elicitation.
\end{abstract}


\maketitle

\section{Introduction} \label{sec:intro}
\begin{figure}
    \centering
    \includegraphics[width=0.8\linewidth]{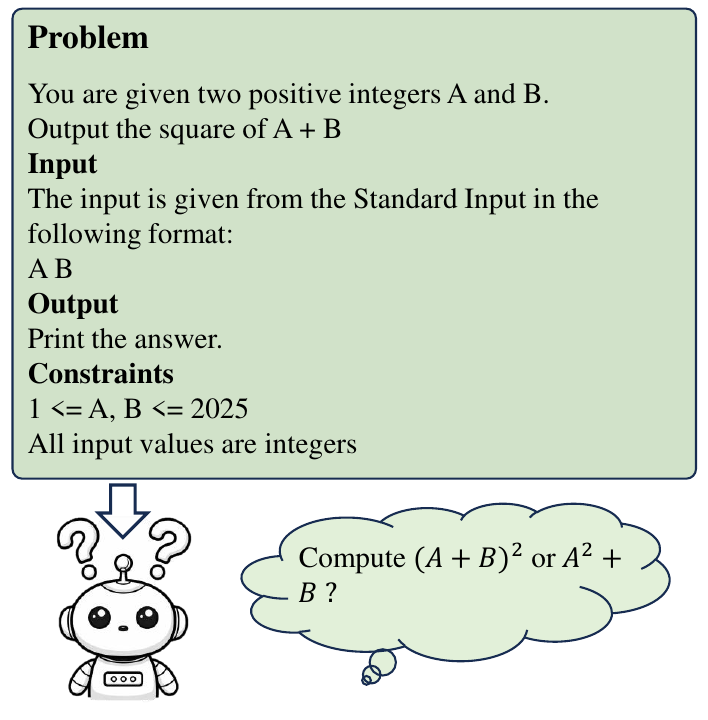}
    \caption{An ambiguous code generation requirement.}
    \Description{An ambiguous code generation requirement that can lead to multiple valid interpretations.}
    \label{fig:a_example}
\end{figure}

Large language models (LLMs) are increasingly used as programming assistants~\cite{cite1, cite2}. Prior studies have reported productivity gains when developers use LLMs for software engineering tasks~\cite{cite3,cite4}. Among these tasks, code generation is one of the most fundamental~\cite{cite5}. In this setting, users provide requirements in natural language, and LLMs generate executable code accordingly~\cite{cite5, cite6}. However, the requirements presented to LLMs are often ambiguous or incomplete~\cite{clarifygpt}. For example, Figure~\ref{fig:a_example} shows a code generation requirement that admits at least two interpretations: whether the target expression is $(A+B)^2$ or $A^2+B$. Such ambiguity can hurt code generation performance and, more importantly, produce outputs that do not match the actual intent of users~\cite{clarifygpt, larbi2025prompts}. This challenge is particularly important in practical development settings, where requirements are rarely fully specified at the beginning of an interaction. This motivates a central question: \textbf{When facing ambiguous requirements, can LLMs recognize the ambiguity and clarify it correctly through interaction with users?}

Existing code generation benchmarks mainly evaluate whether an LLM can generate functionally correct code from a fully specified requirement. Widely used datasets, such as HumanEval~\cite{cite5}, MBPP~\cite{cite6}, and LiveCodeBench~\cite{livecodebench}, usually provide clear and complete problem descriptions. Some recent interactive benchmarks study multi-turn code generation~\cite{convcodeworld, codeif}, but they mainly evaluate the instruction-following ability of LLMs after the requirement has already been refined. They do not examine whether an LLM can first detect ambiguity, ask the appropriate clarification questions, and then use the returned answers to generate correct code. As a result, current benchmarks do not adequately measure the clarification ability of LLMs in realistic coding scenarios, where user requirements are often unclear at the beginning.

\begin{table}[t]
\centering
\small
\setlength{\tabcolsep}{3pt}
\caption{Comparison of code generation benchmarks.}
\label{tab:benchmark_comparison}
\begin{tabular*}{\columnwidth}{@{\extracolsep{\fill}}lcccc@{}}
\toprule
Benchmark & \shortstack{Func.\\Corr.} & Interact. & \shortstack{Ambig.\\Ann.} & \shortstack{Ambig.\\Type} \\
\midrule
HumanEval~\cite{cite5}             & \cmark & \xmark & \xmark & \xmark \\
MBPP~\cite{cite6}                  & \cmark & \xmark & \xmark & \xmark \\
APPS~\cite{apps}                   & \cmark & \xmark & \xmark & \xmark \\
LiveCodeBench~\cite{livecodebench} & \cmark & \xmark & \xmark & \xmark \\
ConvCodeWorld~\cite{convcodeworld} & \cmark & \cmark & \xmark & \xmark \\
CodeFlowBench~\cite{codeflowbench} & \cmark & \cmark & \xmark & \xmark \\
HumanEvalComm~\cite{humanevalcomm} & \cmark & \cmark & \xmark & \textbf{\textit{Coarse}} \\
\textbf{\model}                    & \cmark & \cmark & \cmark & \cmark \\
\bottomrule
\end{tabular*}
\end{table}

To address this gap, we introduce \model, an interactive benchmark for code generation that evaluates how well LLMs clarify ambiguous requirements. Each task in \model{} contains an underspecified requirement, a set of annotated key clarification questions and answers, ambiguity types, and an executable test suite for the final implementation. We evaluate each LLM under an interactive protocol designed to reflect practical code generation workflows. Given an input requirement, the model first decides whether the requirement is ambiguous. When it detects ambiguity, it asks a clarification question and receives an answer from the benchmark environment. It then revisits the requirement with the new information and repeats this process until it decides that the ambiguity has been resolved well enough to produce code. This setup supports end-to-end evaluation of both clarification behavior and downstream code quality. To assess clarification quality more directly, we introduce two new metrics, Turn-discounted Key Question Rate (TKQR) and Optimal Round Adherence (ORA). TKQR rewards models for asking annotated key questions early and penalizes delayed or redundant interaction. ORA measures whether a model uses the appropriate number of interaction rounds, which discourages both premature code generation and unnecessary questioning. Together with functional correctness, these metrics capture both the effectiveness and the efficiency of LLMs as interactive programming assistants.

Building on \model, we systematically evaluate six advanced LLMs under this interactive protocol to examine how well they resolve ambiguity in input requirements. In each interaction, a model must decide whether the requirement is clear enough to generate code or whether it should ask another clarification question before proceeding. Our study leads to three main findings. First, strong code generation performance does not inherently translate to effective requirement clarification. Second, while increased computational "thinking" (e.g., via reasoning models) enhances code correctness, it yields marginal gains in identifying ambiguities. Finally, LLMs' clarification performance degrades sharply as the density of ambiguities increases, revealing a significant bottleneck in handling complex, real-world specifications.


To this end, this paper makes the following contributions:
\begin{itemize}[leftmargin=*, itemsep=2pt, topsep=3pt, parsep=0pt, partopsep=0pt]
\item We introduce \model, a novel interactive benchmark for studying how LLMs clarify ambiguous requirements in code generation. It is built from real code generation tasks and includes annotated key questions, answers, ambiguity types, and executable test suites.
\item We present two new evaluation metrics, TKQR and ORA, to measure both the quality and efficiency of clarification. 
\item We evaluate six advanced LLMs on \model{} and find that current models still struggle to identify and resolve ambiguity. Reasoning does not improve this ability, and performance drops sharply as ambiguity increases.
\end{itemize}

\begin{figure*}[t]
    \centering
    \includegraphics[width=\textwidth]{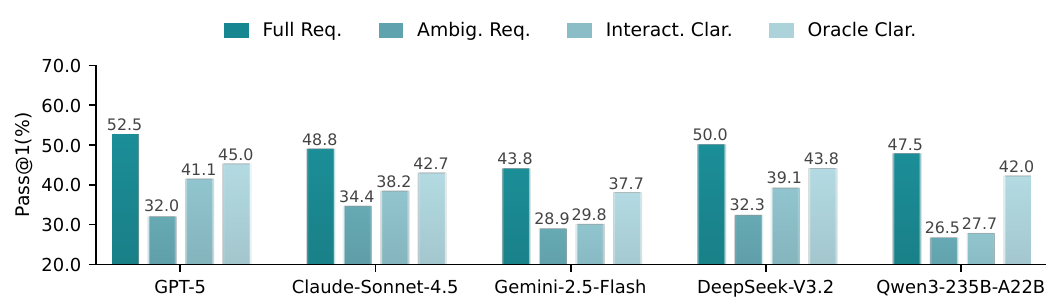}
    \caption{Pass@1 (\%) under four input settings: Complete Requirement (Full Req.), Ambiguous Requirement (Ambig. Req.), Interactive Clarification (Interact. Clar.), and Oracle Clarification (Oracle Clar.).}
    \Description{Pass@1 comparison across five LLMs under four requirement settings in the motivation experiment.}
    \label{fig:motivation}
\end{figure*}
\section{Background} \label{sec:background}
\subsection{Code Generation Benchmarks}

Code generation benchmarks have evolved from simple function synthesis tasks to more realistic settings with richer context. HumanEval~\cite{cite5} and MBPP~\cite{cite6} evaluate models on standalone Python problems with unit tests. APPS~\cite{apps} broadens this setup by introducing thousands of problems across a wider range of difficulty levels. LiveCodeBench~\cite{livecodebench} further improves realism by continuously collecting recent programming problems from online platforms, which helps reduce the risk of contamination. Taken together, these benchmarks provide a solid basis for evaluating code generation across different tasks and settings.

More recent work has introduced interactive benchmarks to study multi-turn coding behavior. ConvCodeWorld~\cite{convcodeworld} examines how models use several forms of feedback, including compilation feedback, execution feedback, and verbal feedback, during iterative code generation. CodeFlowBench~\cite{codeflowbench} focuses on multi-turn coding workflows that more closely resemble real development practice. These benchmarks move beyond single-turn generation and capture more realistic forms of interaction.

Even so, existing benchmarks still assume that the task requirement is sufficiently specified at the outset. Their main purpose is to evaluate whether LLMs can generate correct code, revise code in response to feedback, or follow refinement instructions over multiple turns. They do not directly evaluate whether an LLM can identify ambiguity in an initial requirement, ask appropriate clarification questions, and resolve that ambiguity before generating code. As a result, they offer only limited support for evaluating clarification ability under ambiguous requirements.

HumanEvalComm~\cite{humanevalcomm} is the closest benchmark to our setting. It goes beyond standard code generation evaluation and begins to examine whether LLMs can respond to problematic requirements through clarification. This is an important step toward more realistic evaluation. However, its support for ambiguity analysis remains limited. In particular, it does not provide fine-grained annotations of the specific ambiguity in each instance or the ambiguity type involved. This makes it difficult to measure which kinds of ambiguity are easier or harder for LLMs to identify and clarify. It also limits systematic error analysis across ambiguity categories. Therefore, although HumanEvalComm shows the value of studying clarification, it does not yet support precise evaluation of ambiguity understanding and clarification quality.

Table~\ref{tab:benchmark_comparison} summarizes the differences between existing benchmarks and our dataset. Existing benchmarks mainly evaluate functional correctness or interactive coding behavior. Our dataset, in contrast, focuses on ambiguity-aware clarification and provides explicit ambiguity annotations together with fine-grained ambiguity types.


\subsection{Clarification under Ambiguous Requirements}

Recent studies have begun to examine clarification in code generation. Some work designs interactive workflows that refine requirements before generation. For example, TICODER~\cite{ticoder} helps users formalize intent through tests. ClarifyGPT~\cite{clarifygpt} detects ambiguity, asks clarification questions, and then generates code from the refined requirement. Other work seeks to improve clarification behavior at the model level. ClarifyCoder~\cite{clarifycoder} uses synthetic data and fine-tuning to teach models to recognize ambiguity and ask questions rather than directly guessing user intent. 
These studies show that clarification is valuable for LLM-based code generation. However, they mainly focus on building clarification-enabled systems or measuring end-to-end improvements. They provide limited support for fine-grained ambiguity analysis. In particular, they do not systematically characterize different ambiguity types or support detailed diagnosis of which ambiguities are easy or difficult for LLMs to recognize and clarify. Our work complements this line of research by focusing on ambiguity-centered data construction and evaluation.

\section{Benchmark Construction} \label{sec:benchmark}
In this section, we present the motivation for the construction of the \model{} benchmark, describe the methodology of benchmark construction, define the evaluation metrics, and introduce the interactive evaluation protocol.
\subsection{Motivation Experiment}
Figure~\ref{fig:motivation} provides two clear motivations for building a benchmark to evaluate LLMs' ability to clarify ambiguous requirements. First, compared with Complete Requirement, where the model receives the full specification, Ambiguous Requirement causes a large pass@1 drop for all LLMs. This result shows that ambiguity in the requirement has a strong negative effect on code generation quality and final functional correctness. When key details are missing, LLMs often make implicit assumptions and produce code that does not match the intended behavior. Second, both Interactive Clarification and Oracle Clarification improve over Ambiguous Requirement. This shows that clarifying ambiguity before generation can effectively mitigate the problem. The gain is especially clear in Oracle Clarification, where the model is given the annotated ambiguity questions and answers as dialogue context. These results suggest that requirement clarification is not a marginal step, but a necessary condition for reliable downstream code generation. At the same time, Oracle Clarification still remains slightly below Complete Requirement. This finding is consistent with prior studies~\cite{llmgetlost, interactive}. Even after clarification, the interactive form may still be less effective than presenting the full requirement directly from the start. Overall, these results motivate the need for a dedicated benchmark that can evaluate whether LLMs can detect ambiguity, ask the right clarification questions, and use the answers effectively for code generation.

\subsection{Dataset Construction}

\paragraph{Source Tasks.}
We construct our benchmark \model{} on top of LiveCodeBench~\cite{livecodebench}, a recent code generation benchmark with hidden-test evaluation. We choose LiveCodeBench because its original task statements are sufficiently complete for standard code generation and therefore provide a reliable foundation for controlled ambiguity injection. Each problem includes a natural-language specification, example input--output pairs, and hidden test cases for functional evaluation. This structure provides a strong basis for dataset construction, because ambiguity can be introduced in a controlled way by removing necessary information from an otherwise well-formed requirement. We use LiveCodeBench v6 as the source split.
\begin{table*}[t]
\centering
\small
\setlength{\tabcolsep}{4pt}
\caption{Ambiguity taxonomy used in our dataset. The taxonomy is grounded in requirement quality principles and instantiated for code-generation tasks.}
\label{tab:ambiguity-taxonomy}
\begin{tabularx}{\textwidth}{
  >{\raggedright\arraybackslash}p{2.2cm}
  >{\raggedright\arraybackslash}X
  >{\raggedright\arraybackslash}X
}
\toprule
\textbf{Type} & \textbf{Meaning} & \textbf{Example Clarification Question} \\
\midrule

Terminology
& A domain term, action, or state is undefined, overloaded, or open to multiple interpretations.
& What does each of the symbols \texttt{\#}, \texttt{.}, and \texttt{?} in the grid represent? \\ \\

Behavior
& The required function, objective, or side effect is underspecified, so the intended behavior is not uniquely determined.
& Should this operation modify the original object or return a new one? \\

Edge Cases
& Boundary or exceptional conditions are not specified, leaving behavior unclear for special inputs.
& What should be returned when no valid solution exists? \\

Indices \& Ranges
& Index bases, interval boundaries, or inclusion rules are underspecified.
& Are the endpoints inclusive, and are indices 0-based or 1-based? \\

Ordering \& Atomicity
& Temporal order, simultaneity, or indivisible execution assumptions are unclear.
& Are these updates processed one by one or applied atomically? \\

Output Format
& The required output structure, layout, or presentation rule is missing or unclear.
& Should each answer be printed on a separate line? \\

Comparison Rules
& The comparison key, tie-breaking rule, or stability requirement is not specified.
& How should ties be broken when two items have the same score? \\

Units
& A quantity is specified without a clear unit, scale, prefix, or dimensional convention.
& Should the result be reported in seconds, minutes, or another unit? \\

Collection Semantics
& A collection, container, or state object is mentioned, but its membership, update rule, or access semantics are underspecified.
& What makes an element active, and how should it be retrieved? \\

Numerical Precision
& Precision, rounding, tolerance, or error handling requirements are unclear.
& How many decimal places should be printed? \\

\bottomrule
\end{tabularx}
\end{table*}
\paragraph{Ambiguity Annotation.}
We create ambiguous requirements through \textbf{deletion-only} editing. For each source task, annotators first read the full problem statement and identify the information required to determine the intended behavior. They then remove a small portion of this information to make the requirement ambiguous, while keeping the task natural and preserving the original programming objective. We do not add new content, change the underlying solution, or rewrite the task into a different problem.

For each task, we introduce one to three ambiguity points. For every ambiguity point, we annotate one key clarification question and one ground-truth answer. Each answer is taken directly from the deleted information. Under this design, each edited task is ambiguous initially, but becomes fully specified once the clarification answers are provided.

\paragraph{Quality Control.}
All annotations are conducted by two PhD students. They first annotate the tasks independently. They then cross-check the edits, clarification questions, and answers produced by each other. When disagreements arise, they discuss the case and revise the annotation until consensus is reached. During this process, we explicitly verify three properties: (1) the edited requirement remains natural and plausible, (2) the ambiguity is genuine and cannot be resolved reliably from the remaining context alone, and (3) the annotated question--answer pair is sufficient to recover the original intent. This procedure improves both annotation quality and annotation consistency, while reducing subjective drift across annotators.


\paragraph{Ambiguity Typing.}
\begin{figure}
    \centering
    \includegraphics[width=0.99\linewidth]{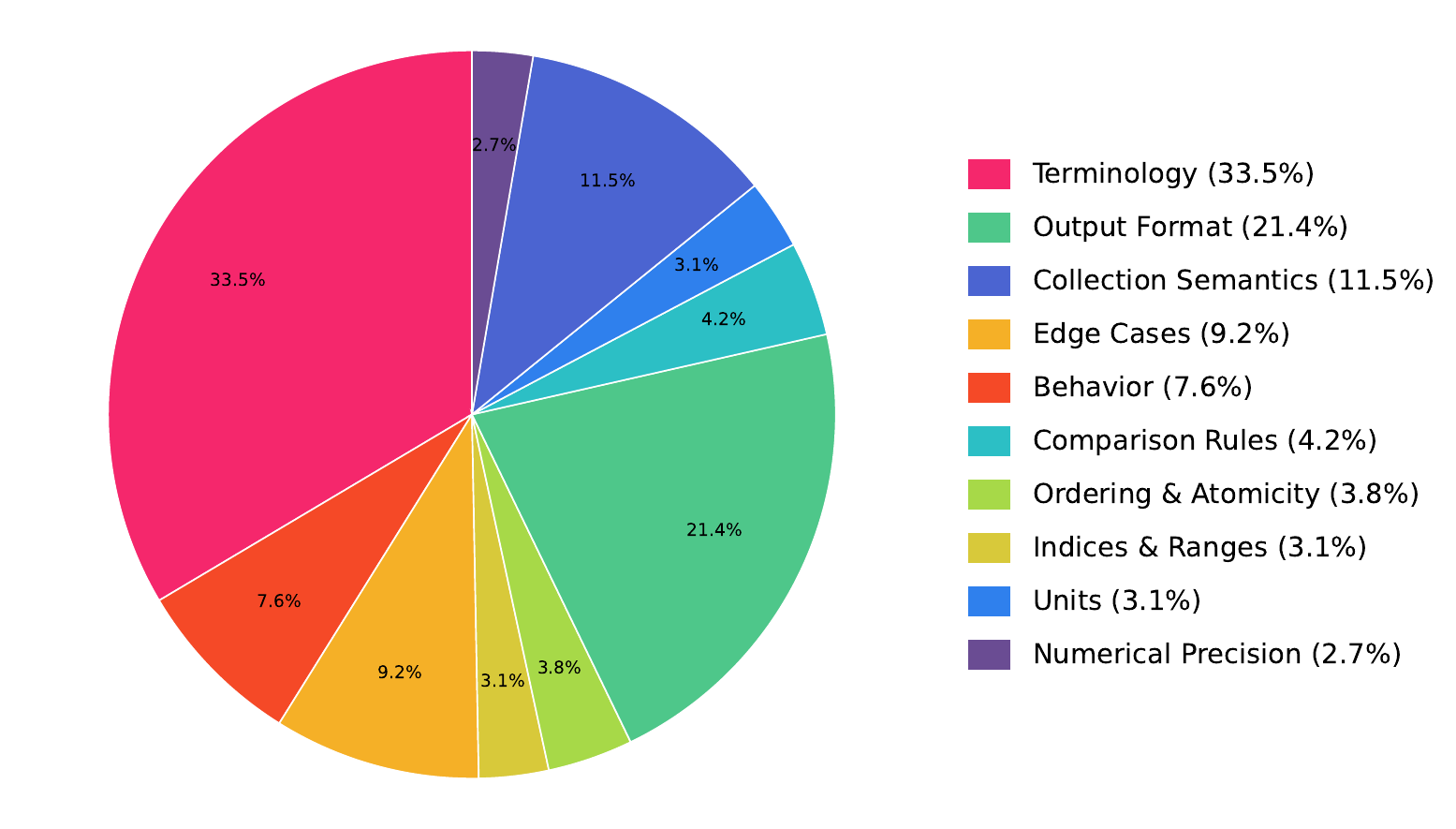}
    \caption{Distribution of ambiguity types in \model{}. \model{} covers a diverse set of ambiguity categories, with Terminology and Output Format being the most common.}
     \Description{Distribution of ambiguity types in the dataset.}
    \label{fig:ambiguity-dist}
\end{figure}
After task construction, we assign a fine-grained ambiguity type to each ambiguity point. The typing scheme is guided by principles of well-formed requirements, especially properties related to clarity, completeness, and verifiability~\cite{iso2018ieee, davis1993identifying}. Figure~\ref{fig:ambiguity-dist} shows the distribution of ambiguity types in the final dataset. Our ambiguity taxonomy is not intended to directly reproduce an external standard. Instead, it is a task-specific taxonomy grounded in requirement quality principles~\cite{davis1993identifying, handbook2003contract}, especially unambiguity, completeness, and verifiability, and instantiated for code generation requirements. Table~\ref{tab:ambiguity-taxonomy} presents the ambiguity taxonomy used in our dataset, together with a brief definition and an example clarification question for each type.

\paragraph{Dataset Summary.}
Following this process, we obtain 419 ambiguous tasks in total. Among them, 199 tasks contain one ambiguity, 169 contain two ambiguities, and 51 contain three ambiguities. Overall, the resulting dataset supports two complementary evaluations: whether an LLM can recognize underspecification and ask the appropriate clarification questions, and whether it can generate correct code after receiving the clarification answers in an interactive setting.

\subsection{Evaluation Metrics} \label{sec:metrics}
To better evaluate the ability of an LLM to handle ambiguous requirements and ask clarification questions, we propose two new metrics: Turn-discounted Key Question Rate (TKQR) and Optimal Round Adherence (ORA). These metrics explicitly capture interaction efficiency, because early identification of missing key information is more valuable, whereas additional turns introduce measurable dialogue cost. We introduce TKQR and ORA below.

\subsubsection{Turn-discounted Key Question Rate}
We define Turn-discounted Key Question Rate (TKQR) by adapting normalized DCG~\cite{jarvelin2002cumulated, jarvelin2017ir} to reward the early asking of key clarification questions. Let $n$ denote the total number of dialogue turns before the model stops asking questions. Let $K$ denote the number of annotated key questions for the task. We define a hit sequence $H=(h_1,\dots,h_n)$, where $h_i\in\{0,1\}$. We set $h_i=1$ if the model asks a previously uncovered key question at turn $i$, so repeated questions receive no additional credit. Otherwise, $h_i=0$. We then compute a discounted gain that favors early hits:
\begin{equation}
\mathrm{DCG}_n \;=\; \sum_{i=1}^{n}\frac{h_i}{\log_2(i+1)}.
\end{equation}
To make scores comparable across tasks with different values of $K$, we normalize by the ideal case. The ideal case asks key questions as early as possible, up to $\min(n,K)$ turns:
\begin{equation}
\mathrm{IDCG}_n \;=\; \sum_{i=1}^{\min(n,K)}\frac{1}{\log_2(i+1)}.
\end{equation}
Finally, TKQR is defined as the normalized ratio:
\begin{equation}
\mathrm{TKQR} \;=\; \frac{\mathrm{DCG}_n}{\mathrm{IDCG}_n}.
\end{equation}
TKQR lies in $[0,1]$. It increases when the model covers key questions earlier, and it decreases when the model delays key questions or spends turns on non-key questions.

\subsubsection{Optimal Round Adherence}
ORA measures whether a model uses a near-optimal number of clarification rounds. We treat the number of question-asking rounds as a cost, because additional rounds increase interaction overhead. Let $n$ denote the number of rounds in which the model asks clarification questions. Let $\mathcal{Q}$ denote the set of annotated key questions for the task. We define the optimal interaction round count as $K = |\mathcal{Q}| + 1$. The additional round accounts for the final step in which the model stops asking questions and transitions to code generation. ORA assigns the highest score when $n=K$, and it decreases as $n$ moves away from $K$. We use a Gaussian-shaped penalty:
\begin{equation}
\mathrm{ORA}(n,K,\sigma)\;=\;\exp\!\left(-\frac{(n-K)^2}{2\sigma^2}\right).
\end{equation}
This form is bounded in $(0,1]$ and decays smoothly with the deviation $|n-K|$. We set $\sigma$ using a simple rule. Specifically, we require $\mathrm{ORA}=0.5$ when $|n-K|=0.5K$. This yields
\begin{equation}
\sigma \;=\; \frac{0.5K}{\sqrt{2\ln 2}} \;\approx\; 0.425K.
\end{equation}
ORA complements TKQR. TKQR evaluates whether key questions appear early, whereas ORA evaluates whether the model stops after a reasonable number of rounds.

\subsection{Evaluation Protocol}





% ...

% ...

\begin{figure}[t]
\centering
\begin{tcolorbox}[
  enhanced,
  width=\columnwidth,
  colback=white,
  colframe=black!55,
  boxrule=0.6pt,
  arc=1.5mm,
  left=0mm,
  right=0mm,
  top=0mm,
  bottom=0mm
]

\PromptBlock{spgray}{
\textbf{You are a professional software developer.}
Your primary goal is to write Python code based on user requirements.
}{System Message}

\PromptBlock{spblue}{
Before coding, you must first evaluate the requirement carefully.

\begin{enumerate}[leftmargin=4mm,itemsep=0.4mm,topsep=0.8mm]
\item If the requirement is clear, write the code directly.
\item If the requirement is unclear, do not guess. Ask exactly one critical clarification question.
\end{enumerate}
}{Task Policy}

\PromptBlock{sporange}{
\textbf{Your response must be in one of the following two formats, with no extra text.}

\vspace{1mm}
{\ttfamily\RaggedRight
[CODE]\par
\{Your code here\}\par
[/CODE]
}

\vspace{1mm}
{\centering\itshape or\par}
\vspace{1mm}

{\ttfamily\RaggedRight
[QUESTION]\par
\{Your single question here\}\par
[/QUESTION]
}
}{Output Format}

\PromptBlock{spgreen}{
{\ttfamily\RaggedRight
[EXAMPLE1]\par
...\par
[CODE]\par
...\par
[/CODE]\par
[/EXAMPLE1]
}

\vspace{1mm}

{\ttfamily\RaggedRight
[EXAMPLE2]\par
...\par
[QUESTION]\par
...\par
[/QUESTION]\par
[/EXAMPLE2]
}
}{Few-shot Demonstration}

\PromptBlock{spblueb}{
Output only one of the two formats above. Do not add any extra explanation or commentary.
}{Output Constraint}

\end{tcolorbox}
\caption{System prompt used in our interactive evaluation. The model must either output code or ask one clarification question.}
\Description{A structured system prompt shown as a stacked template with five colored sections: system message, task policy, output format, few-shot demonstration, and output constraint.}
\label{fig:system-prompt}
\end{figure}

\begin{figure}
    \centering
     \includegraphics[width=0.99\linewidth]{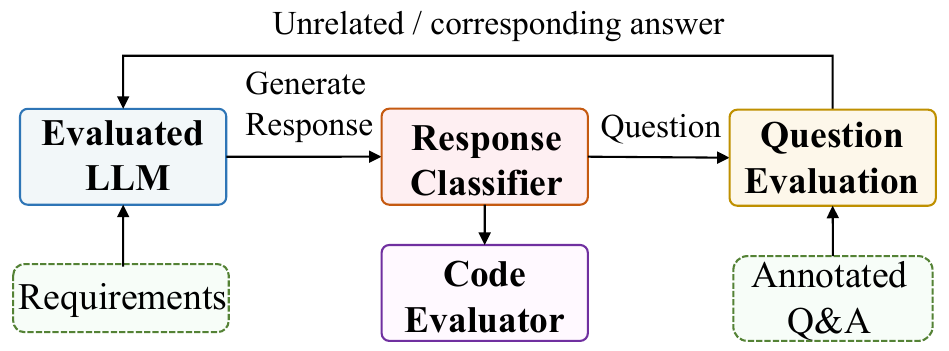}
    \caption{Overview of the evaluation protocol.}
    \Description{..}
    \label{fig:overview}
\end{figure}

\paragraph{Components.}
Figure~\ref{fig:overview} presents the interactive evaluation loop. We first provide the evaluated model with a system prompt, shown in Figure~\ref{fig:system-prompt}, to standardize model behavior across tasks and ensure consistent interactions. Under this prompt, the model must take one of two actions. If it identifies ambiguity in the requirement, it must ask exactly one clarification question. Otherwise, it must directly return the final code.

\paragraph{Multi-round Interaction.}
For each task, we first present the ambiguous requirement to the model and parse its response. If the response is code, the interaction terminates, and we execute the hidden tests to measure functional correctness under the fully specified requirement. If the response is not code, we treat it as a clarification question. We then return an answer and continue the interaction. This process proceeds round by round until the model outputs code or reaches a predefined maximum number of rounds, thereby preventing unbounded interaction.

\paragraph{LLM-as-a-Judge.}
To determine whether a clarification question is valid, we use an LLM-as-a-judge~\cite{zheng2023judging, badshah2025reference} to match the question generated by the model against the annotated key questions for the task. To reduce stochasticity, we repeat the matching process three times and adopt the majority vote. If the question matches an annotated key question, we return the corresponding ground-truth answer to the evaluated model. Otherwise, we return a default reply indicating that no ambiguity exists with respect to that question. This design enables us to evaluate whether the model can identify the correct missing information and request it through interaction, rather than relying on unsupported assumptions.
\section{EXPERIMENTAL SETUP} \label{sec:experiments}
Using \model{}, we conduct a comprehensive study to evaluate the capability of LLMs in code generation under ambiguous requirements, guided by the following research questions.
\begin{itemize}
  \item \textbf{RQ1: (Effectiveness of clarification question asking by LLMs)} How well do recent advanced LLMs perform in asking clarification questions?
  \item \textbf{RQ2: (Performance across ambiguity types)} How do different ambiguity types affect the behavior and performance of LLMs?
  \item \textbf{RQ3: (Performance across difficulty levels)} How do different difficulty levels affect the behavior and performance of LLMs?
  \item \textbf{RQ4: (Clarification depth on \model{})} How many turns of clarification do LLMs attempt on \model{}?
  \item \textbf{RQ5: (Error analysis)} What error patterns do LLMs exhibit on \model{}?
\end{itemize}
\subsection{Studied LLMs}
We evaluate a broad set of advanced LLMs, covering both non-reasoning and reasoning models. The models in our study include GPT-4o~\cite{gpt4o} and GPT-5~\cite{gpt5} from OpenAI, Gemini-2.5-Flash~\cite{comanici2025gemini} from Google, Claude-Sonnet-4.5~\cite{claude-sonnet-4-5-system-card} from Anthropic, DeepSeek-V3.2~\cite{liu2025deepseek} from DeepSeek, and Qwen3-235B-A22B~\cite{yang2025qwen3} from Qwen. These models have achieved strong results on a range of code generation benchmarks, suggesting that they are capable of handling practical code generation tasks.

\subsection{Evaluation Metrics}
We evaluate functional correctness using \textit{pass@1}~\cite{cite5, jiang2024self}. To assess the clarification ability of each model, we use the two metrics introduced in Section~\ref{sec:metrics}: \textit{TKQR} and \textit{ORA}. Together, these metrics provide complementary views of model performance by capturing both downstream code correctness and the quality of clarification behavior.

\subsection{Implementation Details}
We set the maximum number of interaction rounds to 6. For Non Reasoning runs, we limit the output to 1{,}024 tokens. For Reasoning runs, we allow up to 8{,}192 tokens. We use TKQR and ORA to measure the quality of clarification behavior. We set the temperature to 0 for all evaluated LLMs. For GPT-4o~\cite{gpt4o}, which we use as the judge model, we set the temperature to 1.0 and sample three samples, then use majority voting.

\section{RESULTS} \label{sec:results}

\subsection{RQ1: Effectiveness of clarification question asking by LLMs.}


\begin{table}[t]
\centering
\small
\setlength{\tabcolsep}{3pt}
\caption{Clarification quality and code-generation performance. }
\label{tab:main-results}
\resizebox{\columnwidth}{!}{%
\begin{tabular}{@{}l
  S[table-format=1.2]
  S[table-format=1.2]
  S[table-format=2.1,round-mode=places,round-precision=1]
  S[table-format=2.1,round-mode=places,round-precision=1]
@{}}
\toprule
Model
& \multicolumn{1}{c}{TKQR $\uparrow$}
& \multicolumn{1}{c}{ORA $\uparrow$}
& \multicolumn{1}{c}{pass@1 (Ambig.) $\uparrow$}
& \multicolumn{1}{c}{pass@1 (Full)$\uparrow$} \\
\midrule

\multicolumn{5}{l}{\textbf{No-thinking}} \\
\midrule
\texttt{GPT-4o}            & 0.22 & 0.33 & 27.2 & 35.0 \\
\texttt{Claude-Sonnet-4.5} & 0.12 & 0.22 & 38.2 & 48.8 \\
\texttt{Gemini-2.5-Flash}  & 0.29 & 0.23 & 29.8 & 43.8 \\
\texttt{DeepSeek-V3.2}     & 0.30 & 0.50 & 39.1 & 50.0 \\
\texttt{Qwen3-235B-A22B}    & 0.07 & 0.13 & 27.7 & 47.5 \\

\midrule
\multicolumn{5}{l}{\textbf{Thinking}} \\
\midrule
\texttt{Claude-Sonnet-4.5} & 0.12 & 0.20 & 34.3 & 50.0 \\
\texttt{GPT-5}             & 0.21 & 0.26 & 41.1 & 52.5 \\

\bottomrule
\end{tabular}%
}
\end{table}
Table~\ref{tab:main-results} shows that even advanced LLMs struggle to ask useful clarification questions when requirements are ambiguous. Clarification quality is low across all models, both in early coverage of key questions, as measured by TKQR, and in the use of near-optimal interaction rounds, as measured by ORA. DeepSeek-V3.2 achieves the highest TKQR, at 0.30, and the highest ORA, at 0.50, while most other models remain below 0.30 on TKQR and below 0.35 on ORA. This weakness also degrades downstream code generation under ambiguity. For every model, pass@1 (Ambig.), which evaluates \model{} after our ambiguity-resolution pipeline, is substantially lower than pass@1 under complete requirements, with absolute drops ranging from 7.8 to 19.8 points. Taken together, these results indicate that current LLMs often fail to identify missing information, ask the appropriate follow-up question, and recover the actual user intent under ambiguity. As a result, even strong models cannot maintain their code generation performance under complete requirements when missing information must be resolved through interaction.

\finding{1}{Current advanced LLMs have limited ability to ask effective clarification questions under ambiguous requirements. Clarification quality remains low, and this weakness causes large drops in final code generation performance.}

Thinking models can produce stronger code under complete requirements~\cite{li2025structured, guo2025deepseek}, but they do not substantially improve clarification quality. After thinking is enabled, Claude-Sonnet-4.5 increases pass@1 under complete requirements from 48.8 to 50.0, but pass@1 (Ambig.) decreases from 38.2 to 34.3. GPT-5 achieves the strongest overall result in the thinking setting, reaching 52.5 on pass@1 under complete requirements and 41.1 on pass@1 (Ambig.). These results suggest that thinking is beneficial once the requirement has been fully specified. However, this advantage does not extend to clarification. For Claude-Sonnet-4.5, TKQR remains 0.12, and ORA decreases slightly from 0.22 to 0.20. GPT-5 also shows only modest clarification quality, with a TKQR of 0.21 and an ORA of 0.26. This pattern suggests that stronger reasoning does not directly translate into better question asking. When requirements are ambiguous, thinking models may allocate too much of their budget to internal reasoning rather than asking a short and effective clarification question at the appropriate round. Overall, current thinking models appear better at exploiting fully specified requirements than at resolving ambiguity through interaction.

\finding{2}{Thinking helps code generation under complete requirements more than it helps clarification. It can improve raw functional correctness, but it does not reliably improve the ability to ask effective clarification questions under ambiguous requirements.}

\subsection{RQ2: Performance across ambiguity types}
\begin{figure}[t]
    \centering
    \includegraphics[width=0.99\linewidth]{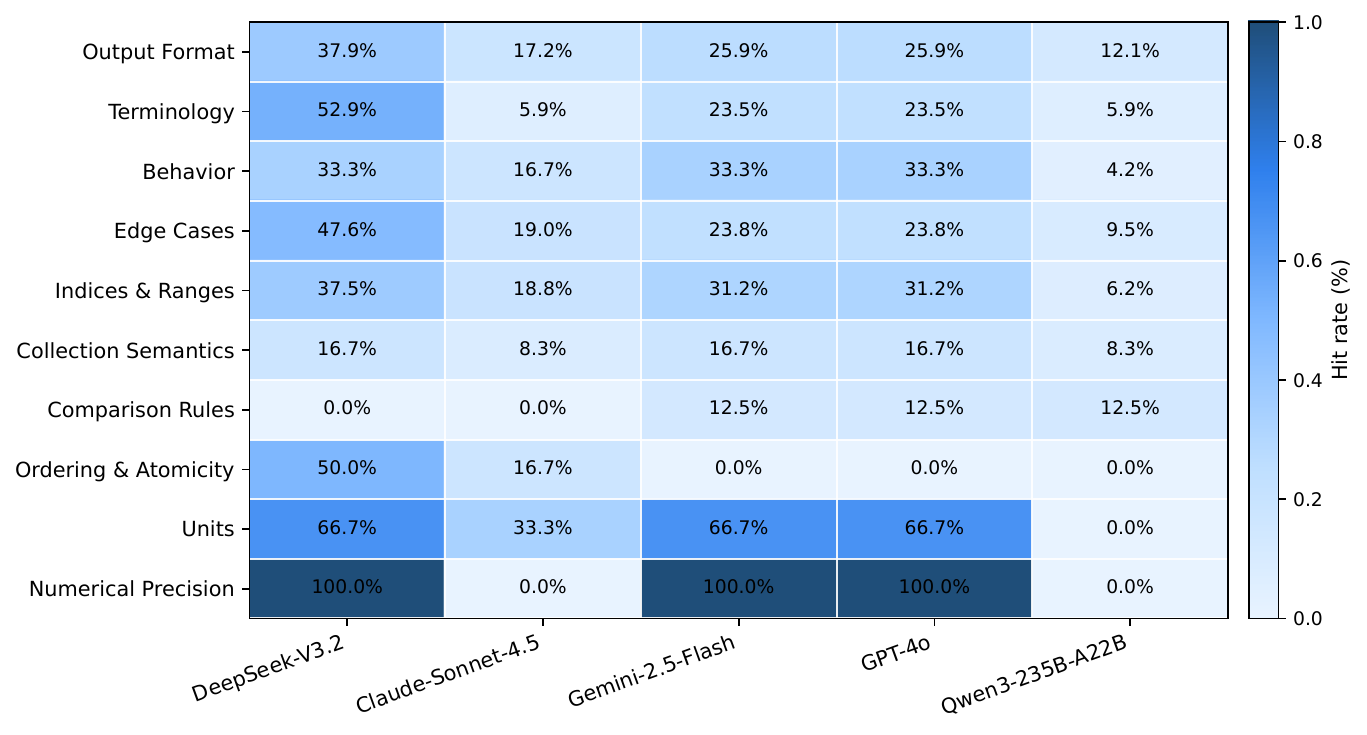}
    \caption{Hit rate across ambiguity types for non-thinking models.A hit means that an LLM asks a clarification question that matches the annotated key question for an ambiguity instance, while a miss means that no such matched clarification question is asked.}
    \Description{Heatmap of hit rates across ambiguity types for five non-thinking models. Rows denote ambiguity types, and columns denote models. Darker cells indicate higher hit rates.}
    \label{fig:types}
\end{figure}

Figure~\ref{fig:types} reveals substantial variation across ambiguity types. Ambiguities involving Units and Numerical Precision are resolved most reliably. DeepSeek-V3.2, Gemini-2.5-Flash, and GPT-4o each reach 66.7\% on Units, and the same three models each reach 100.0\% on Numerical Precision. By contrast, ambiguities that require deeper reasoning about latent semantics remain considerably more difficult. Collection Semantics remains at or below 16.7\% for every model, Comparison Rules reaches only 12.5\% at best, and Ordering \& Atomicity is resolved only by DeepSeek-V3.2 and Claude-Sonnet-4.5, while the other three models remain at 0.0\%. These results suggest that current LLMs are more effective when missing information can be recovered as an explicit local constraint, but they struggle when successful clarification depends on inferring hidden behavioral logic or execution order.

Performance also varies substantially across models. DeepSeek-V3.2 is the most robust model overall and leads on most ambiguity types, particularly Terminology, Edge Cases, and Ordering \& Atomicity. Gemini-2.5-Flash and GPT-4o exhibit a narrower profile of strengths: they remain competitive on Units and Numerical Precision, but are clearly weaker on several semantics-heavy categories. Qwen3-235B-A22B performs worst overall and records 0.0\% on Ordering \& Atomicity, Units, and Numerical Precision. Overall, the results indicate that ambiguity type is a major determinant of clarification difficulty and that current models generalize unevenly across ambiguity types rather than failing in a uniform manner.

\finding{3}{LLMs handle ambiguity types unevenly. They are relatively stronger at resolving explicit local ambiguities such as Units and Numerical Precision, but they struggle with semantics-heavy ambiguities such as Collection Semantics, Comparison Rules, and Ordering \& Atomicity.}

\subsection{RQ3: Performance across difficulty levels}
\begin{figure}
    \centering
     \includegraphics[width=0.99\linewidth]{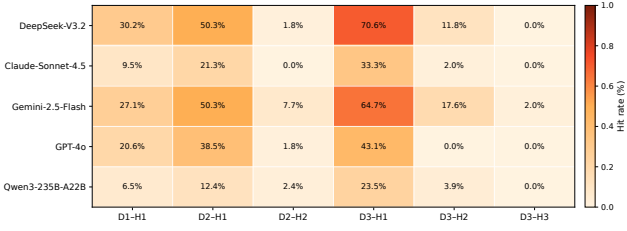}
    \caption{Hit rate across difficulty levels. D$n$-H$m$ indicates that a requirement contains $n$ ambiguities and the LLM successfully asks $m$ matched questions.}
    \Description{}
    \label{fig:diff}
\end{figure}
Figure~\ref{fig:diff} shows that clarification becomes substantially more difficult as the number of ambiguity points in a requirement increases. When only one key clarification question is required, the models achieve limited but still non-trivial success. For example, the best D1-H1 rate is 0.30 for Deepseek-V3.2, followed by 0.27 for Gemini-2.5-Flash. However, performance declines sharply when a task requires multiple key questions to fully resolve the ambiguity. On D2 tasks, models often ask one correct key question, but they rarely ask both. The best D2-H2 rate is only 0.08, achieved by Gemini-2.5-Flash, while several models remain near zero. This pattern is even more pronounced on D3 tasks. Most models can identify one correct key question in some cases, but they rarely continue to uncover all remaining ambiguity points. D3-H2 remains low across all models, and D3-H3 is almost zero for every model. These results suggest that current LLMs can sometimes detect part of the missing information, but they struggle to sustain multi-step clarification when several ambiguity issues must be resolved jointly.

\finding{4}{Clarification difficulty increases sharply with the number of ambiguity points in a requirement. Current LLMs can sometimes identify one missing detail, but they rarely ask all key clarification questions required for requirements with multiple ambiguities.}

\subsection{RQ4: Clarification depth on \model{}}

\begin{table}[t]
\centering
\small
\setlength{\tabcolsep}{4pt}
\caption{Average interaction turns, turn difference, and clarification quality across LLMs. Diff denotes the gap between the actual interaction turns and the ideal number of turns.}
\label{tab:turns}
\resizebox{\columnwidth}{!}{%
\begin{tabular}{@{}l
                S[table-format=1.2]
                S[table-format=+1.2]
                S[table-format=1.2]
                S[table-format=1.2]
                @{}}
\toprule
Model & \multicolumn{1}{c}{Turns} & \multicolumn{1}{c}{Diff} & \multicolumn{1}{c}{TKQR $\uparrow$} & \multicolumn{1}{c}{ORA $\uparrow$} \\
\midrule

\multicolumn{5}{l}{\textbf{No-thinking}} \\
\texttt{GPT-4o}             & 1.26 & -0.38 & 0.22 & 0.33 \\
\texttt{Claude-Sonnet-4.5}  & 0.29 & -1.36 & 0.12 & 0.22 \\
\texttt{Gemini-2.5-Flash}   & 2.22 & +0.57 & 0.29 & 0.23 \\
\texttt{DeepSeek-V3.2}      & 1.54 & -0.10 & 0.30 & 0.50 \\
\texttt{Qwen3-235B-A22B}    & 0.46 & -1.18 & 0.07 & 0.13 \\

\addlinespace[2pt]
\multicolumn{5}{l}{\textbf{Thinking}} \\
\texttt{Claude-Sonnet-4.5}  & 0.24 & -1.41 & 0.12 & 0.20 \\
\texttt{GPT-5}              & 0.70 & -0.94 & 0.21 & 0.26 \\

\bottomrule
\end{tabular}%
}
\end{table}
Table~\ref{tab:turns} reports two complementary statistics. Turns denotes the number of clarification questions that a model asks before generating code, excluding the final round of code generation. Diff denotes the gap between the actual number of clarification rounds and the annotated number of required rounds in the dataset. A negative value indicates that the model asks fewer questions than required.

The results show that current LLMs generally do not clarify ambiguous requirements sufficiently. For most models, the average number of clarification turns is low. This result indicates that they often choose to generate code directly rather than resolve the ambiguity through interaction first. The Diff values further reinforce this pattern. Nearly all models have negative Diff values, indicating that they ask fewer clarification questions than the task actually requires. This gap is substantial for several models, including Claude-Sonnet-4.5 and Qwen3-235B-A22B. Notably, Gemini-2.5-Flash asks more questions, but its low TKQR indicates that many of these questions do not target the key ambiguities. Even stronger models such as DeepSeek-V3.2 still fall short of the required interaction depth on average.

Thinking does not change this overall pattern. Although thinking can improve downstream functional correctness, it does not lead models to ask more clarification questions. For example, the thinking version of Claude-Sonnet-4.5 shows slightly fewer turns and an even larger negative Diff than the no-thinking version. GPT-5 also remains below the annotated interaction depth on average. This pattern suggests that thinking mainly strengthens internal reasoning after a model commits to an interpretation, rather than encouraging it to request missing information from the user. When the requirement is ambiguous, current models still tend to rely on internal inference rather than sustained clarification.

\finding{5}{Current LLMs do not interact deeply enough under ambiguous requirements. They usually ask fewer clarification questions than necessary and often stop the interaction too early. Thinking improves internal reasoning, but it does not reliably increase the depth of clarification.}

\subsection{RQ5: Error Analysis}
We further conduct a manual error analysis to better understand how ambiguity affects code generation. Across the incorrect cases, we identify three major error types: Runtime Error, Wrong Answer, and Time Limit Exceeded. Notably, many of these failures are, in principle, avoidable. Under the original complete requirements, many of the same tasks can be solved correctly. These errors arise because key requirement details are removed, the resulting ambiguity remains unresolved, and the model generates code based on an incorrect interpretation. This result suggests that ambiguity affects not only final accuracy, but also the specific failure mode that emerges in the generated code.
\begin{figure}[t]
\centering
\begin{tcolorbox}[
  enhanced,
  width=\columnwidth,
  colback=white,
  colframe=black!50,
  boxrule=0.6pt,
  arc=1.5mm,
  boxsep=0mm,
  left=0mm,
  right=0mm,
  top=0mm,
  bottom=0mm
]

\begin{reqbox}
\RaggedRight
There is a stack of 100 cards, each labeled with the integer 0.
Process $Q$ queries. Each query is of one of the following:

\begin{itemize}[leftmargin=4mm,itemsep=0.3mm,topsep=0.6mm]
    \item Type 1: Place a card labeled with an integer $x$ on top of the stack.
    \item Type 2: Remove the top card of the stack and output the integer written on that removed card.
    \hl{Under the constraints of this problem, the stack always has at least one card.}
\end{itemize}

Input.

The input is given as $Q$ followed by $Q$ queries. Each query is either \texttt{1 x} or \texttt{2}.

Output.
Let $q$ be the number of type-2 queries. Print $q$ lines.
The $j$-th line should contain the answer to the $j$-th such query.

Constraints.
$1 \leq Q \leq 100$, $1 \leq x \leq 100$. There is at least one query of type 2.
\end{reqbox}

\begin{tcbraster}[
  raster columns=2,
  raster equal height=rows,
  raster column skip=0mm,
  raster row skip=0mm,
  raster left skip=0mm,
  raster right skip=0mm,
  raster before skip=0mm,
  raster after skip=0mm,
  raster valign=top
]
\begin{codebox}[Correct Code]{casegreenhead}{casegreen}
\begin{lstlisting}[style=casepython]
Q = int(input())
stack = [0] * 100

for _ in range(Q):
    query = input().split()
    if query[0] == '1':
        stack.append(int(query[1]))
    else:
        print(stack.pop())
\end{lstlisting}
\end{codebox}%
\begin{codebox}[Incorrect Code]{caseredhead}{casered}
\begin{lstlisting}[style=casepython]
Q = int(input())
stack = []

for _ in range(Q):
    query = input().split()
    if query[0] == '1':
        x = int(query[1])
        stack.append(x)
    else:
        print(stack.pop())
\end{lstlisting}
\end{codebox}
\end{tcbraster}

\end{tcolorbox}
\caption{Runtime Error Case. The highlighted sentence represents the requirement deleted from the full version of the task.}
\Description{A single-column case study figure with a requirement block at the top and two adjacent code blocks below. One sentence in the requirement is highlighted. The left block shows the correct Python implementation, and the right block shows the incorrect Python implementation.}
\label{fig:case-study-stack}
\end{figure}
\paragraph{Runtime Error.} Figure~\ref{fig:case-study-stack} shows a representative failure case under ambiguous requirements. In the original task, the requirement explicitly states, as highlighted in yellow, that under the constraints of this problem, the stack always has at least one card. With this information, the model generates correct code. In the ambiguous version, this highlighted sentence is removed. As a result, the model does not recognize that an important condition is missing, does not ask a clarification question, and directly generates code based on an incorrect interpretation. This leads to a functionally incorrect implementation. This example illustrates a common failure pattern in our benchmark. When a crucial requirement sentence is deleted, current LLMs often fail to detect the resulting ambiguity and instead proceed with a plausible but incorrect assumption. This case further shows why requirement clarification is important. Reliable code generation under ambiguous requirements depends not only on coding ability, but also on the ability of the model to identify missing information and actively seek clarification before implementation.

\paragraph{Wrong Answer.}
\begin{figure}[t]
\centering
\begin{tcolorbox}[
  enhanced,
  width=\columnwidth,
  colback=white,
  colframe=black!50,
  boxrule=0.6pt,
  arc=1.5mm,
  boxsep=0mm,
  left=0mm,
  right=0mm,
  top=0mm,
  bottom=0mm
]

\begin{reqbox}
\RaggedRight
Find one shortest palindrome that has $S$ as its prefix.

Input.

The input is given from Standard Input in the following format:

\texttt{S}

Output.

Print the answer.
If multiple solutions exist, any of them is accepted.

Constraints.

$S$ is a string of length between $1$ and $500000$, inclusive, consisting of uppercase English letters.

\begin{tcolorbox}[
  enhanced,
  colback=casehighlight,
  colframe=casehighlight,
  boxrule=0pt,
  arc=0mm,
  boxsep=0mm,
  left=1.5mm,
  right=1.5mm,
  top=1mm,
  bottom=1mm,
  before skip=1mm,
  after skip=0mm
]
Sample Input 1

\texttt{ABC}

Sample Output 1

\texttt{ABCBA}

ABCBA is a shortest palindrome that has $S=\texttt{ABC}$ as its prefix.
\end{tcolorbox}
\end{reqbox}

\begin{tcbraster}[
  raster columns=2,
  raster equal height=rows,
  raster column skip=0mm,
  raster row skip=0mm,
  raster left skip=0mm,
  raster right skip=0mm,
  raster before skip=0mm,
  raster after skip=0mm,
  raster valign=top
]
\begin{codebox}[Correct Code]{casegreenhead}{casegreen}
\begin{lstlisting}[style=casepython]
S = input()
for i in range(len(S)):
    if S[i:] == S[i:][::-1]:
        print(S + S[:i][::-1])
        break
\end{lstlisting}
\end{codebox}%
\begin{codebox}[Incorrect Code]{caseredhead}{casered}
\begin{lstlisting}[style=casepython]
S = input()
print(S + S[::-1])
\end{lstlisting}
\end{codebox}
\end{tcbraster}

\end{tcolorbox}
\caption{Wrong Answer Case.}
\Description{A single-column case study figure with a requirement block at the top and two adjacent code blocks below. The sample input and sample output section in the requirement is highlighted. The left block shows the correct Python implementation, and the right block shows the incorrect Python implementation.}
\label{fig:case-study-palindrome}
\end{figure}
Figure~\ref{fig:case-study-palindrome} shows a representative \textit{Wrong Answer} case. In the original task, the highlighted sample shows that \texttt{ABC} should produce \texttt{ABCBA}, which makes the shortest-palindrome requirement explicit. In the ambiguous version, this sample is removed. The model does not ask a clarification question and instead assumes that appending the full reverse string is sufficient. This leads to an output that duplicates the middle character and produces a palindrome that is valid but not shortest. This case shows that missing examples can remove key semantic constraints, and current LLMs often fail to recover these constraints without clarification.

\paragraph{Time Limit Exceeded.}
\begin{figure*}[t]
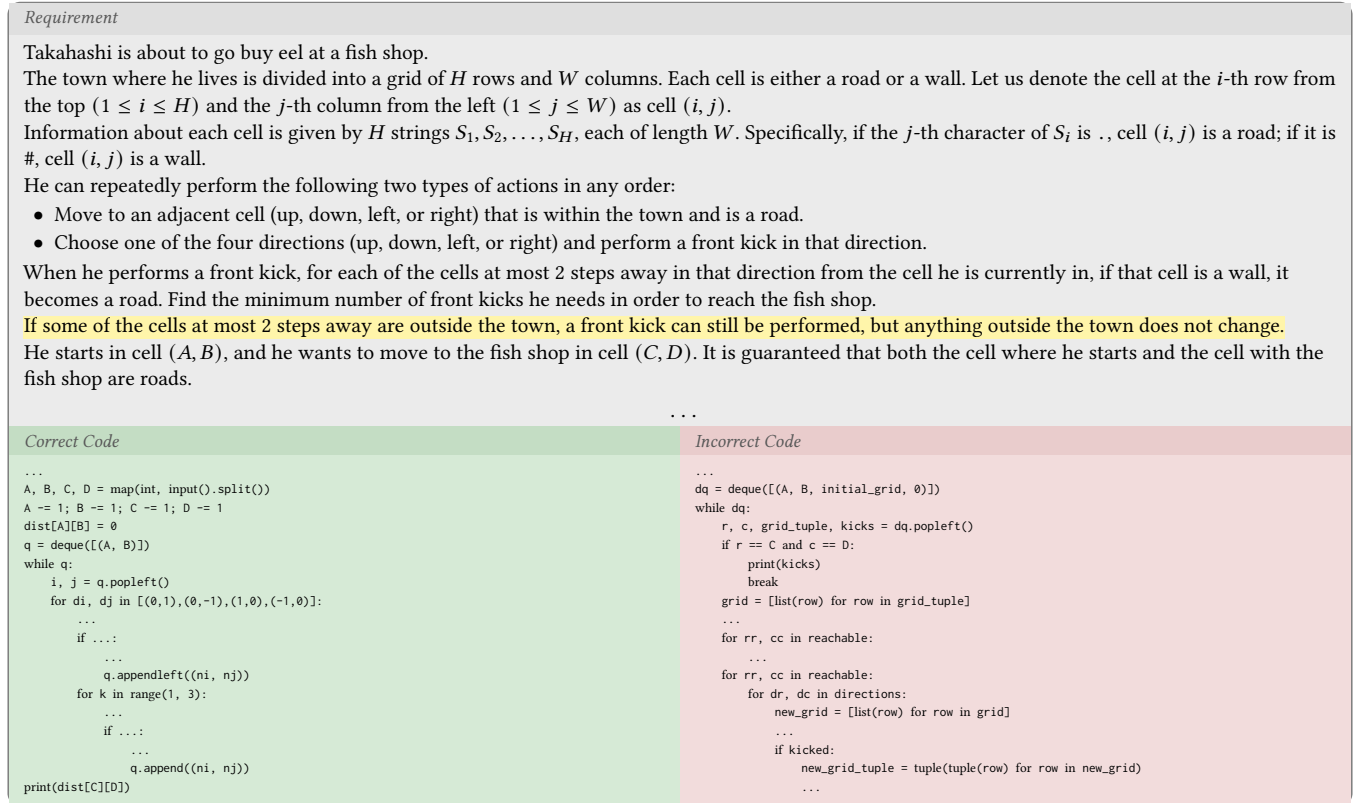

\centering
\begin{tcolorbox}[
  enhanced,
  width=\textwidth,
  colback=white,
  colframe=black!50,
  boxrule=0.6pt,
  arc=1.5mm,
  boxsep=0mm,
  left=0mm,
  right=0mm,
  top=0mm,
  bottom=0mm
]

\begin{reqbox}
\RaggedRight
\small
Takahashi is about to go buy eel at a fish shop.

The town where he lives is divided into a grid of $H$ rows and $W$ columns. Each cell is either a road or a wall. Let us denote the cell at the $i$-th row from the top $(1 \leq i \leq H)$ and the $j$-th column from the left $(1 \leq j \leq W)$ as cell $(i,j)$.

Information about each cell is given by $H$ strings $S_1,S_2,\ldots,S_H$, each of length $W$. Specifically, if the $j$-th character of $S_i$ is \texttt{.}, cell $(i,j)$ is a road; if it is \texttt{\#}, cell $(i,j)$ is a wall.

He can repeatedly perform the following two types of actions in any order:
\begin{itemize}[leftmargin=4mm,itemsep=0.2mm,topsep=0.4mm]
    \item Move to an adjacent cell (up, down, left, or right) that is within the town and is a road.
    \item Choose one of the four directions (up, down, left, or right) and perform a front kick in that direction.
\end{itemize}

When he performs a front kick, for each of the cells at most 2 steps away in that direction from the cell he is currently in, if that cell is a wall, it becomes a road.
Find the minimum number of front kicks he needs in order to reach the fish shop.

\hl{If some of the cells at most 2 steps away are outside the town, a front kick can still be performed, but anything outside the town does not change.}

He starts in cell $(A,B)$, and he wants to move to the fish shop in cell $(C,D)$. It is guaranteed that both the cell where he starts and the cell with the fish shop are roads.

\begin{center}
\texttt{...}
\end{center}
\end{reqbox}

\begin{tcbraster}[
  raster columns=2,
  raster equal height=rows,
  raster column skip=0mm,
  raster row skip=0mm,
  raster left skip=0mm,
  raster right skip=0mm,
  raster before skip=0mm,
  raster after skip=0mm,
  raster valign=top
]
\begin{codebox}[Correct Code]{casegreenhead}{casegreen}
\begin{lstlisting}[style=casepython,basicstyle=\ttfamily\tiny]
...
A, B, C, D = map(int, input().split())
A -= 1; B -= 1; C -= 1; D -= 1
dist[A][B] = 0
q = deque([(A, B)])
while q:
    i, j = q.popleft()
    for di, dj in [(0,1),(0,-1),(1,0),(-1,0)]:
        ...
        if ...:
            ...
            q.appendleft((ni, nj))
        for k in range(1, 3):
            ...
            if ...:
                ...
                q.append((ni, nj))
print(dist[C][D])
\end{lstlisting}
\end{codebox}%
\begin{codebox}[Incorrect Code]{caseredhead}{casered}
\begin{lstlisting}[style=casepython,basicstyle=\ttfamily\tiny]
...
dq = deque([(A, B, initial_grid, 0)])
while dq:
    r, c, grid_tuple, kicks = dq.popleft()
    if r == C and c == D:
        print(kicks)
        break
    grid = [list(row) for row in grid_tuple]
    ...
    for rr, cc in reachable:
        ...
    for rr, cc in reachable:
        for dr, dc in directions:
            new_grid = [list(row) for row in grid]
            ...
            if kicked:
                new_grid_tuple = tuple(tuple(row) for row in new_grid)
                ...
\end{lstlisting}
\end{codebox}
\end{tcbraster}

\end{tcolorbox}
\caption{Time Limit Exceeded Case.}
\Description{A two-column case study figure with a requirement block at the top and two adjacent code blocks below. A highlighted block in the requirement explains the effect of a front kick and the optimization objective. The left block shows the compact skeleton of the correct 0--1 BFS implementation, and the right block shows the compact skeleton of the incorrect implementation that tracks the whole grid state explicitly.}
\label{fig:case-study-eel-kick}
\end{figure*}
Compared with the full requirement, the ambiguous requirement omits the rule that a front kick is still allowed when some cells within two steps are outside the town, and that out-of-bounds cells simply remain unchanged. This omission leaves the kick semantics near the boundary partially underspecified, which can reduce the confidence of the model in applying a higher-level state abstraction. As a result, the same model falls back to a conservative literal simulation under the ambiguous requirement by explicitly maintaining the entire mutable grid, whereas under the full requirement it abstracts the problem as a 0-1 BFS on a static grid. Importantly, the timeout is not caused by ambiguity alone, but by the overly conservative state formulation that unresolved ambiguity makes more likely.

\finding{6}{Ambiguity changes not only whether a model fails, but also how it fails. Once key requirement details are removed and remain unresolved, models often adopt plausible yet incorrect or overly conservative interpretations, leading to avoidable runtime errors, wrong answers, and inefficient implementations that would not appear under the full requirement.}

\section{Discussion} \label{sec:discuss}

\subsection{Agentic Workflows}
Our current evaluation does not include agent-based methods~\cite{xia2025demystifying, dong2024self} because the goal of this study is to isolate model-level clarification behavior under a controlled prompting setting. Recent software engineering agents can interact with tools, inspect repositories, execute tests, and perform multi-step planning~\cite{yang2024swe, zhang2404autocoderover}. These additional capabilities may substantially influence how ambiguity is identified and resolved in real use settings. By contrast, this study examines advanced LLMs under a unified prompting setting in order to preserve comparability across models. Therefore, the findings primarily reflect the clarification ability of base models rather than the end-to-end performance of agent systems. Extending the evaluation to agentic workflows is therefore a natural and important direction for future work.

\subsection{Threats to Validity}
\paragraph{Data Leakage.}
Because our dataset is built from LiveCodeBench problems, data leakage remains a threat to validity. LiveCodeBench was designed to reduce contamination by continuously collecting new problems over time. To further limit this risk, we filtered the source problems and prioritized relatively recent ones during dataset construction. Even so, our ambiguity annotations are added to public problems, and the original problem statements may still have been included in the training data of closed-source LLMs. As a result, some models may infer the intended meaning of an ambiguous requirement from prior exposure rather than from a true ability to ask effective clarification questions. This concern is especially relevant for proprietary models, whose training data is not publicly available. Our results may therefore overestimate the ability of some LLMs to handle ambiguous requirements. Even under this favorable condition, current models still show limited clarification quality, which strengthens our main conclusion that ambiguity remains a major challenge for reliable code generation with LLMs.

\paragraph{LLM-As-Judge.}
A threat to validity arises from our use of an LLM-as-judge to determine whether a clarification question matches the annotated key question. To validate this design, we randomly sampled a subset of instances for human evaluation and compared the human labels with the decisions of the LLM judge. The Cohen's kappa~\cite{cohen1960coefficient, landis1977measurement} between the two sets of labels is 0.89, indicating strong agreement. This result suggests that the judge is highly reliable for this task, although a small amount of residual noise may still remain in borderline cases.


\section{Conclusion} \label{sec:conclusion}
In this paper, we introduced \model{}, an interactive benchmark for evaluating how well LLMs clarify ambiguous requirements before generating code. Using an interactive evaluation protocol and two clarification-oriented metrics, TKQR and ORA, we showed that current LLMs still struggle to identify missing information and ask effective clarification questions. We also found that stronger reasoning does not consistently improve clarification behavior, and that performance drops sharply as ambiguity increases. These findings suggest that reliable LLM-based code generation requires not only stronger synthesis ability, but also better support for ambiguity-aware interaction. We hope that \model{} will support future research on interactive requirement elicitation for AI-assisted software engineering.


\bibliographystyle{ACM-Reference-Format}
\bibliography{Ref}

\end{document}